\documentclass{article}
\usepackage{amsmath,graphicx,spconf}

\title{INERTIA SENSOR AIDED ALIGNMENT FOR BURST PIPELINE IN LOW LIGHT CONDITIONS}
\name{Shuang Zhang, Robert L. Stevenson}
\address{University of Notre Dame, Notre Dame, IN, 46556}
\begin{document}
%
\maketitle
\begin{abstract}
Merging short-exposure frames can provide an image with reduced noise in low light conditions. However, how best to align images before merging is an open problem. To improve the performance of alignment, we propose an inertia-sensor aided algorithm for smartphone burst photography, which takes rotation and out-plane relative movement into account. To calculate homography between frames, a three by three rotation matrix is calculated from gyro data recorded by smartphone inertia sensor and three dimensional translation vector are estimated by matched feature points detected from two frames. The rotation matrix and translations are combined to form the initial guess of homography. An unscented Kalman filter is utilized to provide a more accurate homography estimation. We have tested the algorithm on a variety of different scenes with different camera relative motions. We compare the proposed method to benchmark single-image and multi-image denoising methods with favourable results.
\end{abstract}
\section{Introduction}
\label{sec:intro}
Since the compact structure of a cellphone requires a small-aperture lens, one of the main issues in mobile camera image capture is lack of light. This leads to noisy and poor dynamic range images in low light conditions. Neither increasing gain nor increasing exposure time can attain satisfactory results. The first leads to too much noise and the second to blurry images. Recently, merging a burst of sharp but noisy frames has shown promise. Hasinoff et al. proposed a pipeline to capture a burst of images with a constant short exposure in \cite{Burst:Samuel}. They pick first frame as reference and rest frames as alternatives, then align each alternative to reference along Gaussian pyramids from course to fine and merge aligned frames in frequency domain with a 2D/3D hybrid Wiener filter. This pipeline is proven to almost always outperform conventional single-exposure pipelines in scenes with higher dynamic range, less noise, less motion blur, better color, sharper details, and more texture. 

This pipeline works based on the assumption that only in-plane offset occurs between two frames. However, in-plane rotation and out-plane movement are also common in realistic burst capture. Although within short exposure time and small camera motion, two adjacent frames with small differences can be regarded as in-plane offset, small differences can cumulate to bigger ones, which cannot be canceled by in-plane shift anymore. For example, obvious angles can appear between reference and last alternative. As a result, this assumption limits the number of aligned frame and restrains the capability of denoising.

To remove the in-plane movement limitation, we compute homography using camera-in-build inertia sensor data. Although inertia-sensor is widely used in navigation, simultaneous localization and video stabilization, capturing clear images with inertial sensors using a hand-held camera is more challenging \cite{Joshi:43}. It has been applyed to deblurring problem in \cite{Sindelar:77}\cite{Ruiwen:01}. Since the measurement noise always leads to drift in the estimated motion, generally, these algorithms are either prone to ringing artifacts, or employ time-consuming complex priors. 

Researchers turned to merge a sequence frames with short exposure. Ringaby et al. apply gyro sensors to calculate rotation of camera and the transition motion is estimated from detected feature points by least square \cite{Ringaby:71}. The aligned frames is averaged to obtain a sharp low-noise image. Though they solve the motion drift by rotation interpolation and parameter optimization, it doesn't work well under low-light condition, since not enough points can be captured to calculate  transition. And if there are misaligned frames, merging all frames by averaging will cause ghost effect, which limits the robustness of pipeline. 

In this paper, we address these issues and propose an alignment method aided by smartphone build-in gyroscope. We first calculate initial homography by integrating gyro data. Since the motion drift, this initial homography cannot be directly applied to the current frame. To tackle this issue, we detect matching feature points \cite{Feature} in current and target frames and then plug them into the Unscented Kalman filter. After applying the estimated homography to current frame, an additional Gaussian pyramid based alignment algorithm is adopted to fix possible offset error. To improve the robustness of merge, we use a hybrid Wiener filter in frequency domain and realize a clear result. We apply this pipeline to burst of Bayer raw images and demosaic the merged one into conventional RGB image. The detailed steps are described in section \ref{sec:method}. In section \ref{sec:result}, we demonstrate and discuss the result for three different relative motions: in-plane offset, in-plane rotation and small out-plane movement. The conclusion is made in section \ref{sec:conclusion}.

\section{Method}
\label{sec:method}
\subsection{Compute initial homography by gyro data}
\label{ssec:initialhomo}
Homography is a reasonable way to realize alignment, since it relates to rotation and transition of camera. In short exposure condition, two adjacent frames can be regarded as two planes. And four pairs of points in two frame generates a unique homography. When more than four pairs of points is detected, homography can be obtained by least square method. However, this method only works well when only in-plane rotation and transition happen. The data from camera-build-in inertia sensor is adopted to solve this issue.

While taking pictures with hand-hold smart phone, the corresponding angular velocity,  start time of each exposure and actual exposure time are being recorded. The camera rotation can be recovered from integrating the gyro data. The nonlinear differential equation that relates the rate of change of the rotation matrix to the gyroscope signals is
\begin{equation}
\frac{dR_j}{dt} = \omega_j \times R_j = \left[ 
\begin{array}{ccc} 
0 & -\omega_{jz} & \omega_{jy}\\
\omega_{jz} & 0 & -\omega_{jx}\\
-\omega_{jy} & \omega_{jz} & 0
\end{array}\right] R_j
\end{equation}
where $\omega_j = [\omega_{jx},\omega_{jy},\omega_{jz}]^T $ represents the measured 3-axis angular velocity, $R_j$ is the rotation matrix from the reference coordinate system to the local coordinate system at step $j$, $j = 1,2,...$. Sampling interval $\Delta t$ is decided by camera writing speed. We can obtain the corresponding start and end time of each frame to compute it. With $\Delta t$, the rotation matrix $R_{gyro}$ is computed by 4th order Runge-Kutta method. To transfer rotation matrix into camera coordinate, we have
\begin{equation} 
R_0 = KR_{gyro}K^{-1}
\end{equation} 
where $K$ is the camera intrinsic matrix related to focal length and size of frames.

Speeded Up Robust Features (SURF) \cite{SURF01}, a modified Scale Invariant Feature Transform (SIFT) algorithm, is utilized to detect similar features of two frames and then the homography matrix is generated by matching the image features. Given the homography estimated by SURF points, 3-by-1 transition displacement $T_0$ and normal vector $n_0$, can be computed numerically \cite{Faugeras:01}. Then initial homography becomes
\begin{equation} 
H_0 = R_0+T_0 n_0^T
\end{equation} 

\subsection{Estimate homography by Unscented Kalman filter}
\label{ssec:esthomo}
Since measurement noise always causes the motion drift, the homography computed from gyro sensor is not accurate enough. To obtain a better one, we use Unscented Kalman filter with homography from inertia senor as initial guess and feature points as observation. 

 The Unscented Kalman Filter (UKF), first proposed in \cite{UKF1997}, is proven to be a better nonlinear estimation method compared to the Extended Kalman Filter (EKF). In \cite{UKF2000}, Wan et al. extend the use of the UKF to a broader class of problems, like machine learning. UKF also is applied to estimate homography for multiple plane tracking by Chari et. al in \cite{UKFHomo2010}. They decompose the homography into 10 states to estimate. Different from their work, we directly estimate 8 parameter in homography to simplify computation.

 We regard first frame as the target frame and the frame to transfer as the current frame. The relation between $i$th (i=1,2,...,n) matching points detected in target frame ($\mathbf{x}_i  = [x_i,y_i,1]^T$) and current frame ($\mathbf{x}_i'  = [x_i',y_i',1]^T$) can be written as:
\begin{eqnarray} \label{eq:nonlinearH1}
\mathbf{x}_i' = H\mathbf{x}_i
\end{eqnarray}
where 
\begin{eqnarray}
H = \left[ 
\begin{array}{ccc} 
h_1 & h_2 & h_3\\
h_4 & h_5 & h_6\\
h_7 & h_8 & h_9
\end{array}\right] 
\end{eqnarray} 
is homography. Our goal is to find its estimation $\hat{H}$. Let $\mathbf{h} = [h_1, h_2,h_3,...,h_9]^T$, $\mathbf{x} =[\mathbf{x}_1,\mathbf{x}_2,...\mathbf{x}_n]^T$ and $\mathbf{x}' =[\mathbf{x}_1',\mathbf{x}_2',...\mathbf{x}_n']^T$, then we have the observation function
\begin{eqnarray} 
\mathbf{x}' = G(\mathbf{x},\mathbf{h})
\end{eqnarray} 
Then our aim becomes obtain estimation of vector $\mathbf{h}$, $\hat{\mathbf{h}}$, which makes $\hat{\mathbf{x}}'= G(\mathbf{x},\hat{\mathbf{h}})$ as close as to $\mathbf{x}'$. We now have a nonlinear system, described by a stationary process with identity state transition matrix and a nonlinear measurement model with additive noise individually:
\begin{eqnarray} \label{eq:process}
\mathbf{h}_{k+1} &=& \mathbf{h}_{k}+\mathbf{n}_{\mathbf{h}k}\\ \label{eq:measure}
\mathbf{x}'&=& G(\mathbf{x},\mathbf{h}_{k})+\mathbf{e}_{k}
\end{eqnarray} 
where $\mathbf{n}_{\mathbf{h}k}$ is process Gaussian white noise with covariance matrix $\mathbf{P}$ and $\mathbf{e}_{k} = \hat{\mathbf{x}}'-\mathbf{x}'$ denotes measurement noise with covariance matrix $\mathbf{Q}$. Before we start UKF process, we also need covariance matrix  $\mathbf{P}_{0}$, related to $\mathbf{h}_0$, to initialize. The homogeneous equation (\ref{eq:nonlinearH1}) can be written with vector $\mathbf{h}$ as $\mathbf{A}\mathbf{h} = \mathbf{0}$, with $\mathbf{A}$ the $2n \times 9$ matrix:
\begin{eqnarray} 
\mathbf{A} = \left[ 
\begin{array}{ccccc}
\mathbf{x}_1 & \mathbf{0} & -x_1\hat{x}_1' & -y_1\hat{x}_1' & -\hat{x}_1'\\
\mathbf{0} & \mathbf{x}_1 & -x_1\hat{x}_1' & -y_1\hat{y}_1' & -\hat{y}_1'\\
\mathbf{x}_2 & \mathbf{0} & -x_2\hat{x}_2' & -y_2\hat{x}_2' & -\hat{x}_2'\\
\mathbf{0} & \mathbf{x}_2 & -x_2\hat{y}_2' & -y_2\hat{y}_2' & -\hat{y}_2'\\
\vdots & \vdots & \vdots & \vdots & \vdots \\
\mathbf{x}_n & \mathbf{0} & -x_n\hat{x}_n' & -y_n\hat{x}_n' & -\hat{x}_n'\\
\mathbf{0} & \mathbf{x}_n & -x_n\hat{y}_n' & -y_n\hat{y}_n' & -\hat{y}_n'
\end{array}
\right]
\end{eqnarray} 
The traditional way such as \cite[p. 142-147]{Hartley:01} requires inverse of $ \mathbf{A}^T \mathbf{A}$ to compute $\mathbf{P}_{0}$, which can be poorly conditioned when correspondences are almost noise-free, like SURF points in our case. Criminisi et. al instead use matrix perturbation theory to compute $\mathbf{P}_{0}$ in \cite{Criminisi:1999}. Assuming the feature points are perturbed with Gaussian noise, we can obtain $\mathbf{P}_{0}$ from matrix $\mathbf{A}$ and covariance of feature points (refer \cite{Criminisi:1999} for details). 

For each step of UKF, we first need to calculate sigma points $\mathcal{X}$, a minimal set of carefully chosen sample points to represent a Gaussian distribution of states. When propagated through the true nonlinear system, in our case $G(\mathbf{x}, \mathbf{h})$, these points capture the posterior mean and covariance accurately. 

For $k$th step, the sigma points with dimension $L\times(2L+1)$ and corresponding weights ($1\times(2L+1)$)have the form
\begin{eqnarray}  \nonumber
\mathcal{X}_{k-1}& = &\left[ \hat{\mathbf{h}}_{k-1}, \hat{\mathbf{h}}_{k-1}\pm\sqrt{(L+\lambda) \mathbf{P}_{k-1}} \right]\\ \nonumber
W^{(m)} &=& \left[ \lambda/(L+\lambda), \lambda/2(L+\lambda), \lambda/2(L+\lambda), \ldots \right]\\ \nonumber
W^{(c)} &=& \left[ \lambda/(L+\lambda)+(1-\alpha^2+\beta),  \lambda/2(L+\lambda), \ldots \right] \nonumber
\end{eqnarray} 
where $L$ is the dimension of state $\mathbf{h}$, $\lambda = (\alpha^2-1)L$ denotes the scaling parameter with $\alpha$ a small number (we choose $\alpha=1e-3$ according to \cite{UKF2000}) and $\beta=2$ is used to incorporate prior knowledge of the Gaussian distribution \cite{UKF2000}. 
To forecast the next step, we have
\begin{eqnarray}  \nonumber
\hat{\mathbf{h}}_k^- &=& \sum\nolimits_{j=0}^{2L} W^{(m)}_j\mathcal{X}_{k-1,j}\\ \nonumber
\mathbf{P}_{k}^- &=&  \sum\nolimits_{j=0}^{2L} W^{(c)}_j(\mathcal{X}_{k-1,j}-\hat{\mathbf{h}}_{k-1}^-)(\mathcal{X}_{k-1,j}-\hat{\mathbf{h}}_{k-1}^-)^T
\end{eqnarray} 
For $j$the column of sigma points, let $\mathcal{Y}_{k-1,j} = G(\mathbf{x},\mathcal{X}_{k-1,j})$, then the mean and covariance for $\mathcal{Y}$ are represented by
\begin{eqnarray}  \nonumber
\hat{\mathbf{y}}_k^- &=& \sum\nolimits_{j=0}^{2L} W^{(m)}_j\mathcal{Y}_{k-1,j}\\ \nonumber
\mathbf{P}_{\mathbf{y}_k\mathbf{y}_k} &=&  \sum\nolimits_{j=0}^{2L}  W^{(c)}_j(\mathcal{Y}_{k-1,j}-\hat{\mathbf{y}}_k^-)(\mathcal{Y}_{k-1,j}-\hat{\mathbf{y}}_k^-)^T
\end{eqnarray} 
To update the state, we have
\begin{eqnarray}  \nonumber
\mathbf{P}_{\mathbf{h}_k\mathbf{y}_k} &=&  \sum\nolimits_{j=0}^{2L}  W^{(c)}_j(\mathcal{X}_{k-1,j}-\hat{\mathbf{h}}_k^-)(\mathcal{Y}_{k-1,j}-\hat{\mathbf{y}}_k^-)^T\\ \nonumber
\mathcal{K} &=& \mathbf{P}_{\mathbf{h}_k\mathbf{y}_k}\mathbf{P}_{\mathbf{h}_k\mathbf{y}_k}^{-1}\\ \nonumber
\hat{\mathbf{h}}_k &=& \hat{\mathbf{h}}_k^-+\mathcal{K} (\mathbf{y}_k-\hat{\mathbf{y}}_k^- )\\ \nonumber
\mathbf{P}_{k}&=&\mathbf{P}_{k}^- -\mathcal{K}\mathbf{P}_{\mathbf{y}_k\mathbf{y}_k}\mathcal{K}^T
\end{eqnarray} 
By UKF, we can obtain a more accurate homography for current frame.  However, there still can be some frames failing to align with the target frame, in most cases last several frames. They should be excluded to ensure successful merging.

\begin{figure}[!t]
	\centering 
	\includegraphics[width=3.2in]{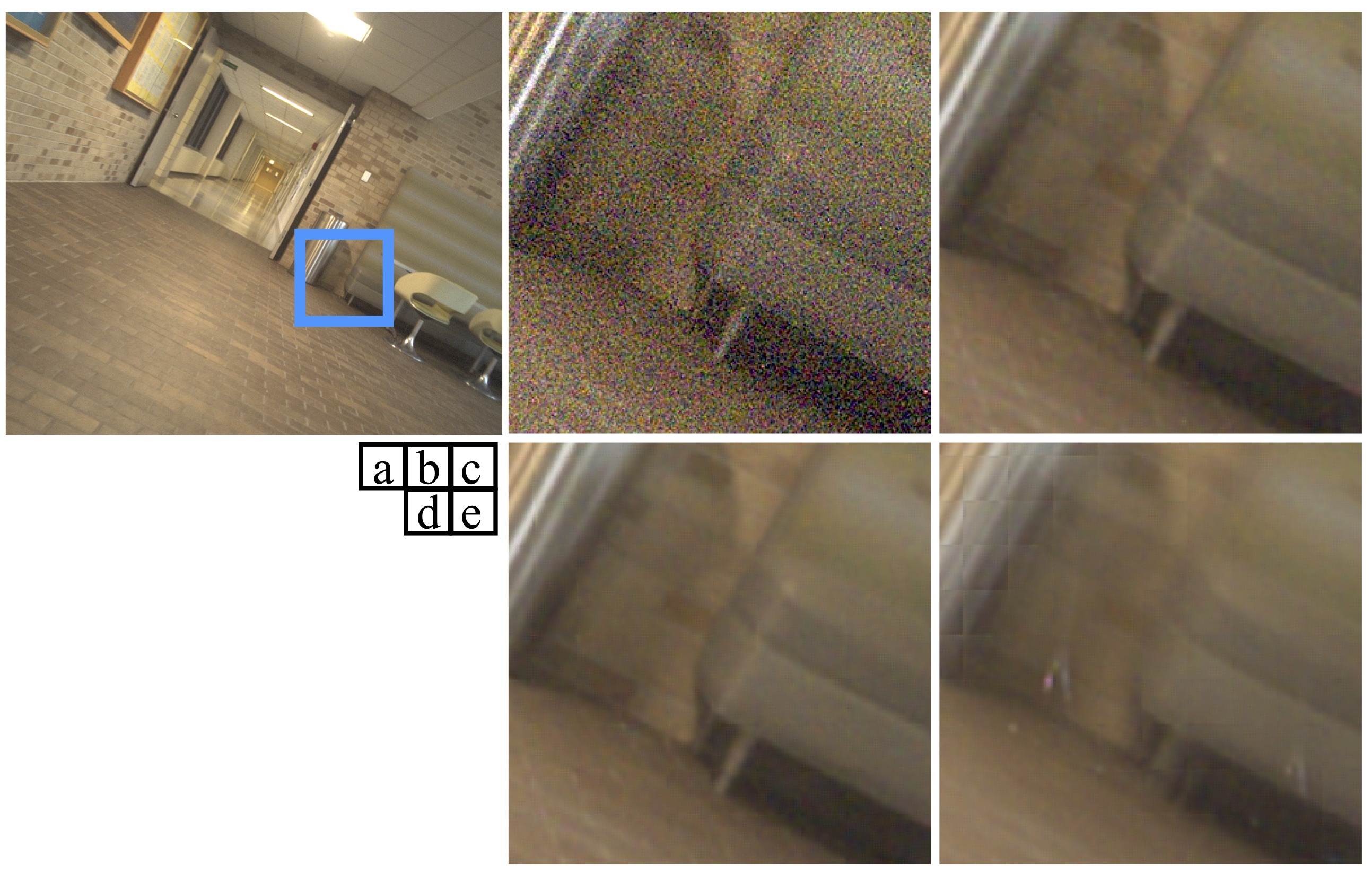} 
	\caption{Denoised results for different number of aligned frames. (a) a full frame denoised after inertia sensor aided method with N=15. (b)(c)(d) are denoised local images with number of frames: 1, 5, 10, 15} \label{fig:num}
\end{figure}

\begin{figure*}[!t]
	\centering 
	\includegraphics[width=6.5in]{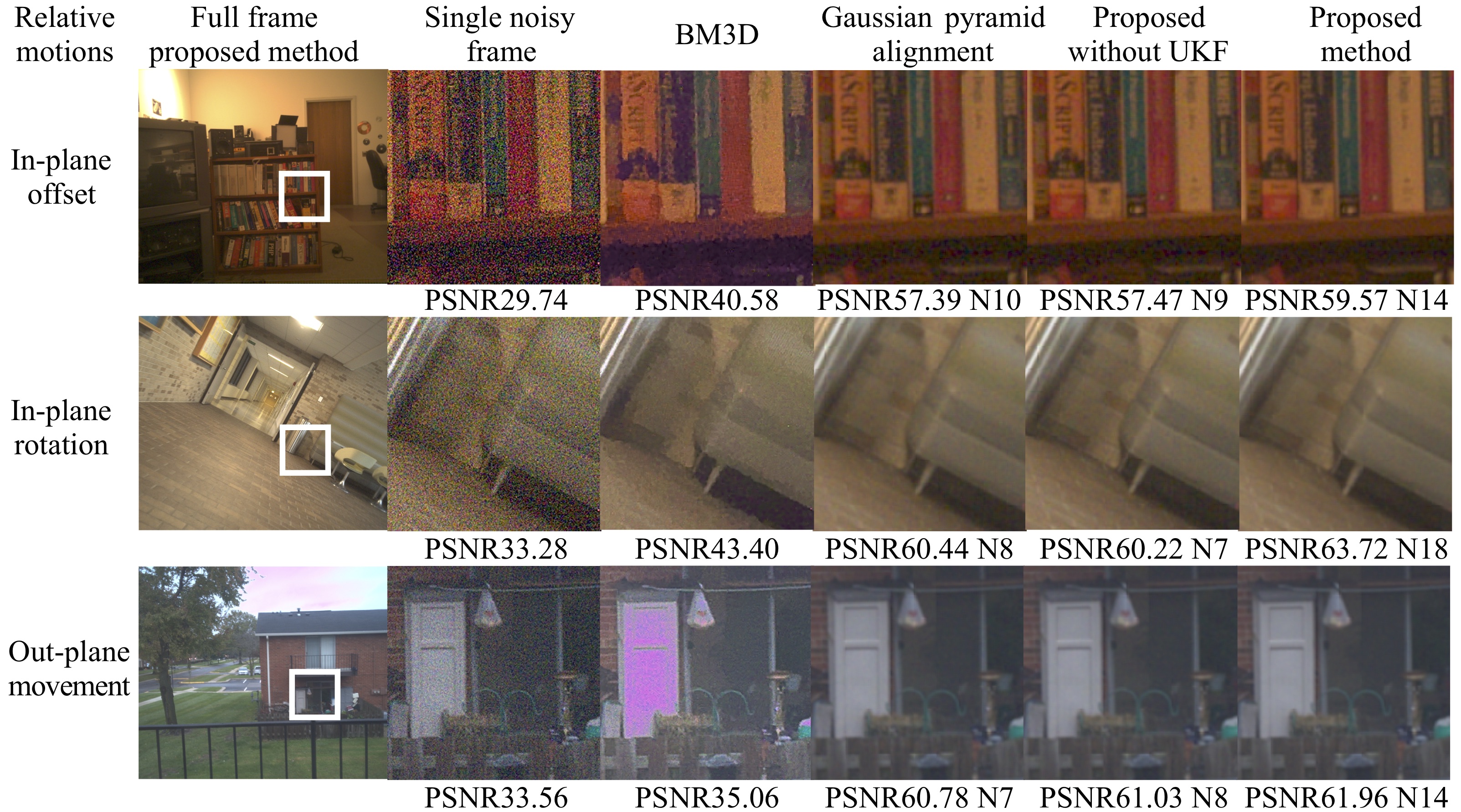} 
	\caption{Denoised results for different relative camera motions. Top to bottom: three different motions of camera when taking pictures. Left to right: full frame which is merged after inertia sensor aided alignment method, first noisy local frame for each input, local result of BM3D \cite{BM3D:CFA}, local result of original pipeline with Gaussian Pyramid alignment method \cite{Burst:Samuel}, local result of feature points without inertia sensor or UKF, local result of proposed alignment method. N is number of valid frames. PSNR in dB. } \label{fig:3Movements}
\end{figure*}
\subsection{Select number of valid frames}
\label{ssec:framenum}
According to \cite{Burst:Samuel}, the variance of noise after merge is $\sigma^2/N$, where $\sigma^2$ is estimated variance of noise in original frames, $N$ is number of aligned frames. It suggests that the more aligned frames are combined, the more noise can be decreased. Hence, the first rule for number of frames is to adopt as many as we can, when a few or no artifacts appear. In experiment, we find that if expectation of steady error $E[\mathbf{e}]=E[\hat{\mathbf{x}}'-\mathbf{x}’]$ between reference and aligned alternative is larger than 5 the result of merge is not desired. In this case, we exclude frames with steady error $E[\mathbf{e}]$ larger than 5. At last, improvements beyond 18 frames are difficult to notice \cite{Ringaby:71}. Fig.\ref{fig:num} shows denoised results for different number of aligned frames. Fig.\ref{fig:num}(b) and (c) includes too few frames; (e) contains too many frames since artifacts appear; (d) is desired one with suitable number of frames.

\section{Result and discussion}
\label{sec:result}
In this part, we process the low-light burst of Bayer raw frames captured by a hand-held camera, LG Nexus 5 (Android 5.0.1, API 21). After stacking frames by inertia sensor data and UKF, we merge valid aligned frames according to a 2D/3D hybrid Wiener filter \cite{Burst:Samuel}. The merged Bayer raw image is converted to RGB image by demosaicking. To provide a desired image, several postprocessing steps are added to merged frame, including color space correction and gamma correction.

We test our approach on real-life sequences with in-plane offset, in-plane rotation and out-plane movement as input relative camera motions. Then we compare our results with the Block Matching 3D denoising method (BM3D) \cite{BM3D:CFA}, Gaussian pyramid burst pipeline \cite{Burst:Samuel} and feature points based alignment without inertia sensor or UKF. Comparison with BM3D can provide us the denoising effect between best conventional single-image (according to \cite{BenchMark}) and novel multi-images denoising methods. Comparison with Gaussian pyramid pipeline shows results between multi-frames denoising algorithms with different alignment methods. The other comparison provides the performance of with/without inertia sensor data and UKF in proposed method.

Since we provide denosing results based on real camera motion and noisy photographs, it is difficult to quantify the denoising effect. To solve this problem, we adopt the single image noise level estimation in \cite{NoiseLevel:2013} and transfer it into Peak signal-to-noise ratio (PSNR) for the processed RGB image.

As shown in Fig. \ref{fig:3Movements}, in general, multi-frame methods achieve clearer results since multiple frames can provide more information about scenes. Even if some details are concealed by noise in one frame, it can be recovered from others. At the same time, multi-frame methods have less artifact of color. For multi-frame methods, the number of successful aligned frames is highly related to noise level as we mentioned. And our method can obtain more aligned frames in all three cases. It means more details can be seen, especially in lower light conditions. For in-plane offset, the proposed method has similar result as Gaussian pyramid alignment method, which means our method conserves the image quality of original alignment method. For in-plane rotation and out-plane movement (rotation around x-axis in this case),our approach adopts more valid frames to merge, which suggests less noise. 

One reason for frame number limitation is that the drift error from inertia sensors cumulates with increasing frames. Since we align each alternative frame to the first one, the first several frames have small differences with the reference; after several seconds, bigger differences appear and are beyond capability of alignment although our method has higher capability. The limitation of inertia sensor accuracy also can be a possible reason.

\section{Conclusion}
\label{sec:conclusion}
In this paper, we introduced an inertia sensor aided algorithm to improve stacking performance of a burst pipeline. We first find initial homography by using gyroscope sensor and the SURF registration algorithm. Then we adopt unscented Kalman filter to improve the accuracy of estimated homography. This algorithm conserves advantages of pipeline in \cite{Burst:Samuel}: robust alignment with constant exposure and Bayer raw frames with less tone-mapping. And more importantly, our method is less sensitive to camera movement and increases the number of valid frames. As a result, it can achieve less noisy results. Even if, under low-light condition, the feature points are difficult to capture, the following Gaussian pyramid alignment can fix small offset caused by lacking of translation information. 



\bibliographystyle{IEEEbib}
\bibliography{strings,refs}

\end{document}